\documentclass[twocolumn]{aastex63}

\usepackage{afterpage}
\usepackage{amsmath}

\submitjournal{ApJ}

\shorttitle{H$_2$O$_2$ greenhouse warming on oxidized early Mars}
\shortauthors{Ito et al.}

\begin{document}

\title{H$_2$O$_2$ induced greenhouse warming on oxidized early Mars}
\footnote{\today}

\correspondingauthor{Yuichi Ito}
\email{yuichi.ito.kkyr@gmail.com}

\author[0000-0002-0598-3021]{Yuichi Ito}
\affiliation{Department of Physics and Astronomy, University College London, London WC1E 6BT, United Kingdom}
\affiliation{Department of Cosmosciences, Hokkaido University, Sapporo 060-0810, Japan}

\author[0000-0002-3821-6881]{George L. Hashimoto}
\affiliation{Department of Earth Sciences, Okayama University, Okayama 700-8530, Japan}

\author[0000-0003-4060-7379]{Yoshiyuki O. Takahashi}
\affiliation{Department of Planetology, Kobe University, Kobe 657-8501, Japan}

\author[0000-0001-7490-4676]{Masaki Ishiwatari}
\affiliation{Department of Cosmosciences, Hokkaido University, Sapporo 060-0810, Japan}

\author[0000-0002-6757-8064]{Kiyoshi Kuramoto}
\affiliation{Department of Cosmosciences, Hokkaido University, Sapporo 060-0810, Japan}

\begin{abstract}
The existence of liquid water within an oxidized environment on early Mars has
been inferred by the Mn-rich rocks found during recent explorations on Mars.
The oxidized atmosphere implied by the Mn-rich rocks would basically be comprised of CO$_2$ and H$_2$O
without any reduced greenhouse gases such as H$_2$ and CH$_4$. So far, however, it has been thought that early Mars could not have been warm enough to sustain water in liquid form without the presence of reduced greenhouse gases.
Here, we propose that H$_2$O$_2$ could have been the gas responsible for warming the surface of the oxidized early Mars.
Our one-dimensional atmospheric model shows that only 1 ppm of H$_2$O$_2$ is enough to warm the planetary surface because of its strong absorption at far-infrared wavelengths, in which the surface temperature could have reached over 273~K for a CO$_2$ atmosphere with a pressure of 3~bar.
A wet and oxidized atmosphere is expected to maintain sufficient quantities of H$_2$O$_2$ gas in its upper atmosphere due to its rapid photochemical production in slow condensation conditions.
Our results demonstrate that a warm and wet environment could have been maintained on an oxidized early Mars, thereby suggesting that there may be connections between its ancient atmospheric redox state and  possible aqueous environment.  
\end{abstract}

\section{Introduction} \label{sec:intro}
One of the most intriguing and debatable problems in planetary science is elucidating how an early Martian surface environment could have been warm enough to sustain liquid water \citep{Wordsworth2016,Ramirez+2018}.
Climate models have shown that a CO$_2$--H$_2$O atmosphere alone could not have kept early Mars warm enough to sustain liquid water globally even if any amount of atmospheric pressure is assumed \citep[e.g.,][]{Kasting1991}. 
This suggests that the other greenhouse components could have played a key role in the early Martian atmosphere.
Current theoretical models suggest that the warming of early Mars was caused by
a CO$_2$--H$_2$O atmosphere combined with additional greenhouse substances; clouds \citep[e.g.,][]{Forget+1997,Wordsworth+2013}; reducing gases such as H$_2$, CH$_4$, and NH$_3$  
\citep[e.g.,][]{Ramirez+2014,Ramirez2017,Wordsworth+2017,Sagan+1972,Kasting+1992}; and/or volcanic gases such as H$_2$S and SO$_2$ \citep[e.g.,][]{Postawko+1986,Johnson+2008,Tian+2010}. 

Recently, \emph{NASA}'s Curiosity rover discovered a high abundance of Mn in sedimentary rocks \citep{Lanza+2016}. 
During the era in which the observed Mn-oxide-rich rocks at Gale crater would have precipitated out, Mars may have had both liquid water on its surface and a highly oxidized atmosphere \citep{Lanza+2016,Noda+2019}.
Furthermore, the existence of rocks with a high concentration of manganese at Endeavour crater \citep{Arvidson+2016} suggests that such an oxidized and wet surface environment was a global phenomenon at that time.
These findings suggest that the early Martian surface had once experienced a wet and warm environment, but with the absence of reduced gas species that would have enhanced the greenhouse effect of a CO$_2$--H$_2$O dominated Martian atmosphere to allow the existence of liquid water. 
One might consider SO$_2$ as a candidate greenhouse gas in an oxidized atmosphere \citep[e.g.,][]{Johnson+2008}, but its presence seems unlikely during this era because Mn and S are not correlated in the rocks found at Gale crater \citep{Lanza+2016}.

\begin{figure}[t]
\begin{center}
\includegraphics[width=\columnwidth]{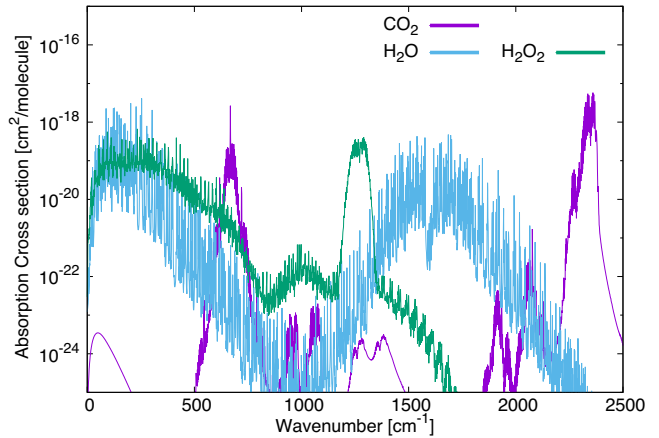}
\\
\caption{
Absorption cross sections of three oxidized gases, CO$_2$ (magenta), H$_2$O (cyan), and H$_2$O$_2$ (green) at 250~K and 1~bar, as functions of wavenumber. These cross sections are produced using the line profile calculation code EXOCROSS \citep{Yurchenko+2018}. The absorption data and the assumed line profiles are described in Sec.~\ref{sec:mod}.
 }
\label{fig:cross}
\end{center}
\end{figure}

\begin{figure}[t]
  \begin{center}
    \includegraphics[width=\columnwidth]{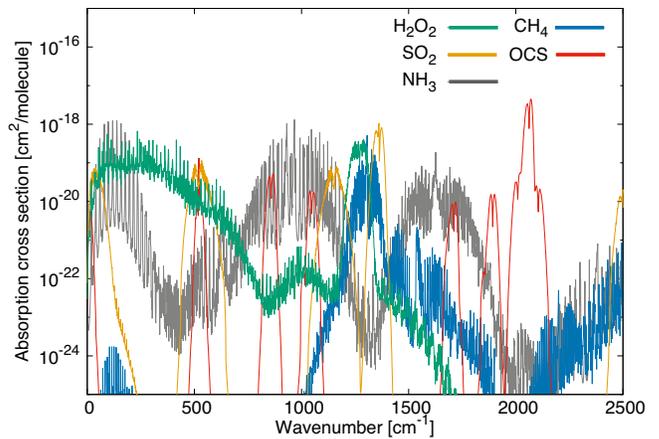}
    \\
    \caption{
Absorption cross sections of H$_2$O$_2$ (green) and four greenhouse gases, SO$_2$ (orange), NH$_3$ (\replaced{yellow}{gray}), CH$_4$ (blue) and OCS (red) at 250~K and 1~bar, as functions of wavenumber. 
These cross sections are calculated using EXOCROSS \citep{Yurchenko+2018} and HITRAN2012 \citep{Rothman+2013}, assuming a Voigt profile truncated at 25 cm$^{-1}$ from the line center.
   }
    \label{fig:cross2}
  \end{center}
\end{figure}

In an attempt to address this uncertainty, in this study we investigate the greenhouse effect due to hydrogen peroxide (H$_2$O$_2$) gas in the early Martian atmosphere.
Previously, it had been proposed that H$_2$O$_2$ gas was responsible for oxidizing the early Martian surface \citep[e.g.,][]{Zahnle+2008}.
Although the idea that H$_2$O$_2$ was one of the greenhouse gases responsible for the warming of early Mars has widely been ignored, H$_2$O$_2$ does absorb radiation at wavenumbers near 500 cm$^{-1}$, where the blackbody radiation at a temperature of $250$~K has peak intensity and CO$_2$ has an absorption window, as shown in Fig.~\ref{fig:cross} \citep[see also Figure~4 in][]{Wordsworth2016}.
Also, the absorption cross section of H$_2$O$_2$ is larger than those of known greenhouse gases such as SO$_2$, NH$_3$, CH$_4$ and OCS in a wavenumber range from 250~cm$^{-1}$ to 450~cm$^{-1}$, as shown in Fig.~\ref{fig:cross2}.

The remainder of this paper is organized as follows. 
In Section~\ref{sec:mod}, we describe our atmospheric model and numerical setup.
In Section~\ref{sec:res} we show the surface temperature as a function of H$_2$O$_2$ abundance under the conditions that may have been present on early Mars.  
We discuss the photochemical production and condensation of H$_2$O$_2$ in a warm and wet early Martian atmosphere and the possible warming scenario of H$_2$O$_2$ in an oxidized early Martian environment in Section~\ref{sec:disc}. Finally we summarize our results in Section~\ref{sec:conc}.

\section{Atmospheric Model} \label{sec:mod}
We set up a vertical, one-dimensional CO$_2$-dominant atmospheric model and determine the surface temperature required to achieve balance between the absorbed solar radiation and the outgoing planetary radiation with approximated radiative--convective equilibrium temperature-pressure profiles and given compositions.
The numerical scheme is based on the same line-by-line calculations used in the calculations of the surface temperature warmed by H$_2$O  \citep{Schaefer+2016} and CO$_2$--H$_2$--CH$_4$ atmospheres \citep{Wordsworth+2017}.

The atmosphere in hydrostatic equilibrium is vertically divided into 100
layers from the ground to the top of the atmosphere  ($1\times10^{-4}$~bar for the model atmosphere).
The surface pressure ranges from 0.01--3 bar.
The vertical grid of the atmosphere is set so that the logarithms of pressure are evenly spaced.  
Following previous models \citep{Ramirez+2014,Ramirez2017}, we set the modeled atmosphere to one composed of 95\% CO$_2$, fully-saturated H$_2$O, fully-saturated or different mixing ratios of H$_2$O$_2$, and $\leq$ 5~\% N$_2$.
For the saturated H$_2$O$_2$ amount, we calculate the vapor pressure using the Clausius-Clapeyron equation. Then, we use the thermal properties of H$_2$O$_2$ \citep{Foley+1951a} and its saturation vapor pressure of $4.69\times10^{-4}$ bar at the melting point (272.69~K) as a reference pressure \citep{Manatt+2004}. 
In the other cases, the molar fraction of H$_2$O$_2$ is assumed to be vertically constant and its value was in the range from 10~ppb to 10~ppm.
Additionally, the abundance of H$_2$O is determined by the saturation vapor pressure of water  \citep[Eqs. 11 and 12 in][]{Kasting+1984}.

The atmospheric temperature profile is assumed to be that of 
a moist adiabat of H$_2$O and CO$_2$ from the surface to the tropopause and an isothermal stratosphere above the tropopause.
Using the heat capacity, $c_p$,  given by the Shomate equation\footnote{\url{http://old.vscht.cz/fch/cz/pomucky/fchab/Shomate.html}}, the vapor amount and latent heat of H$_2$O \citep[Eq.12 in][]{Kasting+1984} and the gravity of Mars, $g$, the moist adiabat of H$_2$O is given by Eq. (2.48) in \citet[][]{Andrews2000}.
Decreasing gravity with altitude is included in this model.
The moist adiabat of CO$_2$ is adopted where the moist adiabat of H$_2$O is colder than the saturation vapor pressure of CO$_2$  using  Eqs. A5 and A6 in \citet[][]{Kasting1991}.
We assume that the stratospheric temperature is 155~K ($\sim167$~K$\times0.75^{1/4}$), which is based on the results of \citet{Kasting1991}, who uses 167~K as the stratospheric temperature for current solar heating and scales it for different solar heating rates by assuming that the stratospheric temperature is proportional to the skin temperature.  

Using the atmospheric structure described above, we calculate the outgoing planetary radiation based on a line-by-line radiative transfer calculation.
The outgoing planetary radiation is given by 
\begin{eqnarray}
F_p&=&    2\pi \int B_\nu(T_{\rm{surf}}) \int^1_0  \mu e^{-{\tau_{\nu,\rm{surf}}}/\mu}d\mu d\nu  \nonumber \\
&&+2\pi \int \int^{\tau_{\nu,\rm{surf}}}_0 \int^1_0 B_\nu(t_\nu)e^{-\tau_\nu/\mu} d\mu d\tau_\nu  d\nu,
 \label{eq:fp}
\end{eqnarray}
where $T_{\rm{surf}}$ is the surface temperature, $\mu$ is the cosine of the zenith angle, $B_\nu$ is the Planck function at wavenumber, $\nu$, and $\tau_{\nu,\rm{surf}}$ is the total optical depth of the atmosphere.
In hydrostatic equilibrium, the optical depth is given by
\begin{equation}
\frac{d\tau_{\nu}}{dP}=\frac{\sum \chi_A \sigma_{\nu,A}}{\bar{m}g},
\end{equation}
where $P$ is atmospheric pressure, $\bar{m}$ is the mean mass of the atmospheric gas particles, and $\chi_A$ and $\sigma_{\nu,A}$ are the molar fraction and absorption cross section of an absorber $A$, respectively.
Following \citet{Ramirez2017} and \citet{Kopparapu+2013}, the line absorption cross section profile of CO$_2$ is assumed to be a sub-Lorentzian \citep{Perrin+1989} truncated at 500~cm$^{-1}$ from the line center, while that of H$_2$O is assumed to be a Voigt profile truncated at 25~cm$^{-1}$ from the line center. 
The line profile of H$_2$O$_2$ is also assumed to be a Voigt profile truncated at 25~cm$^{-1}$ from the line center.
For the line absorption of each gas species, the line data given by HITRAN2012 \citep{Rothman+2013} and the line profile calculation code EXOCROSS \citep{Yurchenko+2018} are used in this model.
Additionally, the collision-induced absorption of CO$_2$--CO$_2$ \citep{Gruszka+1997,Baranov+2004} is considered.
In practice, to save memory and CPU time, we have prepared a numerical table in which the absorption cross sections are given as functions of temperature, $T$, and log$_{10}$ $P$.
The table was created using values of $T=150, 200, 250, 300$ and 350~K, and log$_{10}$ ($P$/bar)$=-4, -3, -2, -1, 0$ and 1.
We evaluate the integral shown in Eq. (\ref{eq:fp}) over a wavenumber range from 1~cm$^{-1}$ to 10000~cm$^{-1}$ with a resolution of 1~cm$^{-1}$.
The numerical integration of Eq.~(\ref{eq:fp}) with respect to the zenith angle is performed using the exponential integral calculation code presented by \citet{Press+1996}, while the other integrals are evaluated using trapezoidal integration. 

We iteratively determine the surface temperature at which the outgoing planetary radiation balances the absorbed solar radiation.  
The absorbed solar radiation is given by $(1-A_p)F_{\rm{sol}}/4$, where $A_p$ is the planetary albedo and $F_{\rm{sol}}$ is the solar flux. 
The solar flux is assumed to be $F_{\rm{sol}}=590\times0.75~$ W/m$^2$, and we use the planetary albedo of a wet CO$_2$(95\%)-N$_2$(5\%) atmosphere not warmed by any additional greenhouse mechanism \citep[][private communication]{Ramirez+2014}.
Note that our assumed planetary albedo underestimates the surface temperature in a warm atmosphere with enhanced saturated--H$_2$O content more than the atmosphere not warmed by H$_2$O$_2$.
This is because the absorption of solar radiation by H$_2$O decreases the planetary albedo \citep{Kasting1988}, and H$_2$O$_2$ might work in the same way.
While the Rayleigh scattering cross section per a H$_2$O$_2$ molecule is comparable with that of CO$_2$ based on its electric dipole polarizability \citep{George+1992}, the abundance of H$_2$O$_2$ in our model is too low (up to 10~ppm) to increase the planetary albedo. Also, though there is no public absorption data of H$_2$O$_2$ in the optical regime
 \deleted{\citep{Jonathan+2018}}
 \added{\citep[see][]{Al-Refaie+2016,Jonathan+2018}},
  its absorption in the optical is likely to not be very strong \citep[see also the MPI--Mainz UV/VIS Spectral Atlas\footnote{\url{http://satellite.mpic.de/spectral_atlas/index.html}};][]{Keller-Rudek+2013}.

\subsection{MODEL VALIDATION} \label{ssec:mv}

\begin{figure}[t]
  \begin{center}
    \includegraphics[width=\columnwidth]{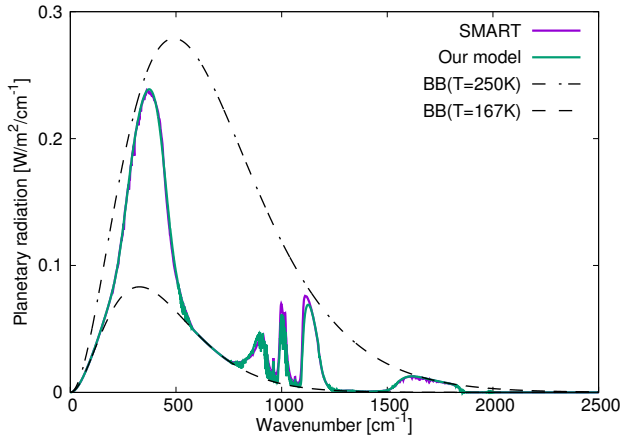}
    \\
    \caption{
    Outgoing planetary radiation as a function of wavenumber for a dry, 2~bar CO$_2$ atmosphere, comparing the result calculated by our line-by-line model (green) against the result produced with the SMARTS code (violet). 
    In each model, a dry CO$_2$(95\%)-N$_2$(5\%) atmosphere with a pressure of 2~bar, a surface temperature of 250~K and a stratospheric temperature of 167~K is assumed. Also, the assumed temperature profile follows the dry and moist adiabatic lapse rate of CO$_2$. The plotted SMART data are the same with those shown in Figure S2 of \citet{Ramirez+2014} (Ramses Ramirez, private communication).
     The black dashed curves show blackbody radiation, of which the temperatures are indicated by BB($T$). 
   }
    \label{fig:comp_1}
  \end{center}
\end{figure}

\begin{figure}[t]
  \begin{center}
    \includegraphics[width=\columnwidth]{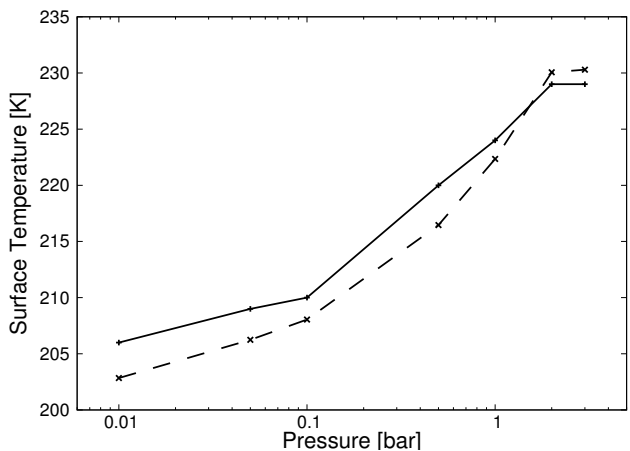}
    \\
    \caption{
Surface temperature as a function of surface pressure for wet CO$_2$(95\%)-N$_2$(5\%) atmospheres, comparing the result calculated by our model (solid) against the result of \citet{Ramirez+2014} (dotted). 
   }
    \label{fig:comp_2}
  \end{center}
\end{figure}

We have performed two benchmark tests of our simulation code, and we have confirmed that our model reproduce the numerical solutions for the dry and wet CO$_2$-rich atmospheres of early Mars shown in \citet{Ramirez+2014}.

We first compare our line-by-line model against a well-tested line-by-line model, SMART \citep{Meadows+1996}, for a dry, 2-bar CO$_2$(95\%)-N$_2$(5\%) atmosphere.
With the same temperature profile shown in Figure S1 of \citet{Ramirez+2014}, we calculated the outgoing planetary radiation using our model. Fig.~\ref{fig:comp_1} shows a comparison between the SMART result and that from our model. Our model spectra agree well with the SMART spectra, although there is some difference in the wavenumber region from 800~cm$^{-1}$ to 1200~cm$^{-1}$, which is likely due to the different absorption data used in both studies.
The total flux of our model is 87.2~W/m$^2$, which agrees well with that found by SMART (88.4~W/m$^2$).
Note that the calculated fluxes differ by at most 0.05~\%, even if we double the resolution of the wavenumber or the number of vertical layers.

Next, we compare the surface temperatures of a wet, 2-bar CO$_2$(95\%)-N$_2$(5\%) atmosphere with that calculated by the one-dimensional radiative convective model \citet{Ramirez+2014}.
Our results agrees well with those of \citet{Ramirez+2014}, as shown in Fig.~\ref{fig:comp_2},
where the differences in the calculated surface temperatures are no more than 4~K.
Because the results of our models agree to within 2~\% of the previous studies, we have confirmed that our model is consistent with these models.

\section{Results} \label{sec:res}
\begin{figure}[t] 
  \begin{minipage}{\columnwidth}
    ($a$) Outgoing planetary radiation
    \begin{center}
      \includegraphics[width=\columnwidth]{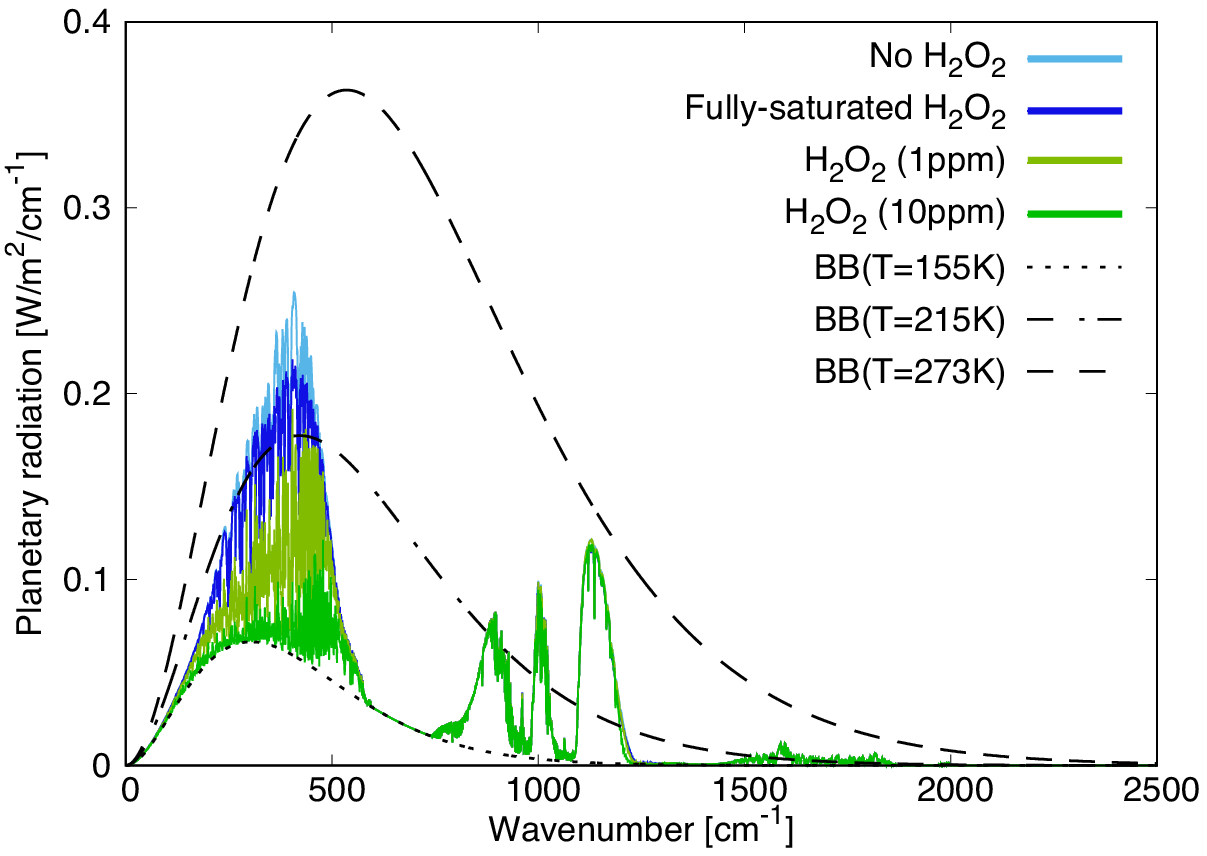}
    \end{center}
  \end{minipage} \\
  \begin{minipage}{\columnwidth}
    ($b$) Atmospheric structure
    \begin{center}
      \includegraphics[width=\columnwidth]{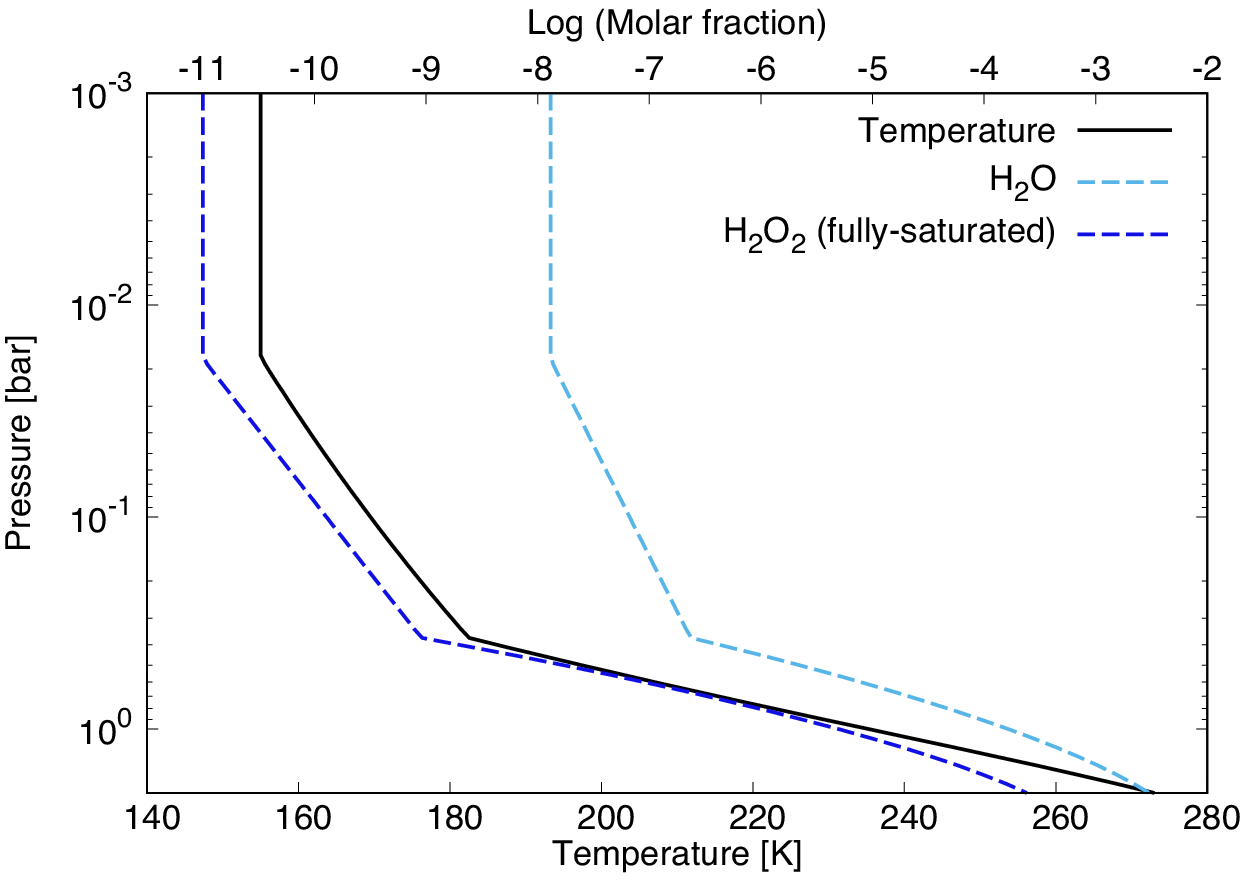}
    \end{center}
  \end{minipage} 
  \caption{
Impact of H$_2$O$_2$ on the outgoing planetary radiation for a surface pressure of 2~bar and a surface temperature of 273~K.
\textit{Panel ($a$)} --- the outgoing planetary radiation as a function of wavenumber for the atmospheres without H$_2$O$_2$ and with different amounts of H$_2$O$_2$. The black dashed curves show the blackbody radiation, the temperatures of which are indicated, BB($T$).
\textit{Panel ($b$)} ---the temperature-pressure profile and the vertical distributions of H$_2$O and saturated H$_2$O$_2$ in the atmosphere.
}
  \label{fig:tpfp}
\end{figure}
 
\begin{figure}[t]
  \begin{center}
    \includegraphics[width=\columnwidth]{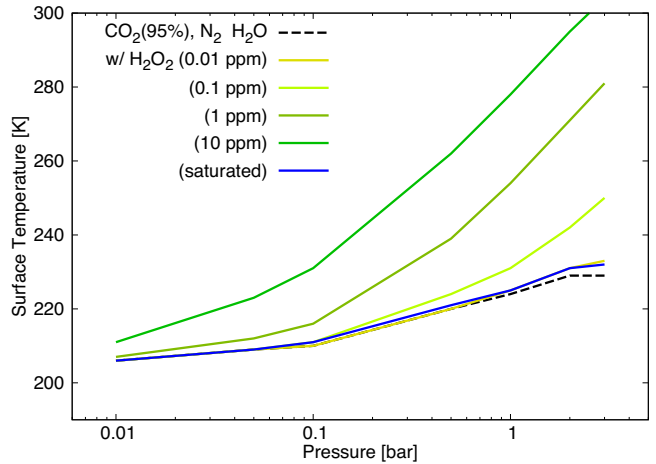}
    \\
    \caption{
Surface temperature as a function of surface pressure.
The dashed curve represents the atmosphere without H$_2$O$_2$ while the solid curves show the atmospheres containing a saturated amount and vertically constant molar fractions of H$_2$O$_2$.
Note that, the case of 0.01 ppm of H$_2$O$_2$ (yellow) is almost identical to those of saturated- (blue) and free-H$_2$O$_2$ (black).  
   }
    \label{fig:h2o2}
  \end{center}
\end{figure}

Fig.~\ref{fig:tpfp}~(a) shows the outgoing planetary radiation for a fixed surface pressure and temperature of 2~bar and 273~K, respectively.
When the atmosphere consists of H$_2$O and CO$_2$ (cyan), there are atmospheric windows at wavenumbers below 500~cm$^{-1}$ and around 1000~cm$^{-1}$, which are consistent with the results of previous climate models \citep[e.g.,][]{Wordsworth2016,Ramirez2017}. 
The addition of H$_2$O$_2$ reduces the planetary radiation at wavenumbers below 500~cm$^{-1}$ due to its strong far--IR absorption (blue, olive, green).
Although H$_2$O$_2$ effectively absorbs photons with wavenumbers around 1200~cm$^{-1}$(Fig.~\ref{fig:cross}), this only slightly affects the outgoing planetary radiation because CO$_2$ also absorbs photons at the same wavenumbers.  

The outgoing planetary radiation is 87.6~W/m$^2$ for an H$_2$O$_2$ free atmosphere (cyan), which decreases drastically when H$_2$O$_2$ is added.  
For vertically constant molar fractions of 1 ppm (olive) and 10 ppm (green) of H$_2$O$_2$, the outgoing planetary radiation is 68.8~W/m$^2$ and 56.5~W/m$^2$, respectively.  
If the abundance of H$_2$O$_2$ can be constrained by the saturation vapor pressure, the planetary radiation is 84.2~W/m$^2$ (blue), and the greenhouse effect of H$_2$O$_2$ is not remarkable.  
This is because the abundance of saturated H$_2$O$_2$ is too low in the low pressure region to absorb photons effectively (Fig.~\ref{fig:tpfp}~(b)).

Next, Fig.\ref{fig:h2o2} shows the surface temperature as a function of surface pressure.  
The differences in surface temperatures between the atmospheres without H$_2$O$_2$ (black) and with saturated H$_2$O$_2$ (blue) are at most 4~K.
However, in the case of abundant H$_2$O$_2$, the planetary surface is warm enough to sustain liquid water (Fig.\ref{fig:h2o2}).
In particular, for the 2~bar atmosphere with added 1~ppm (olive) or 10~ppm (green) of H$_2$O$_2$, the surface temperature increases by about 40~K or 65~K from that of the H$_2$O$_2$-free case ($\sim$ 230~K), respectively.
Our results show that a concentration of only 1 ppm level of H$_2$O$_2$ is sufficient to effectively cut off the outgoing planetary radiation and warm the planetary surface to temperatures above 273~K.  
\added{Note that, these H$_2$O$_2$ are supersaturated at high altitudes. For example, in the 2~bar atmosphere,
1~ppm of H$_2$O$_2$ is supersaturated below the pressure level of $\sim$~0.8~bar, and its supersaturation ratio is as high as ~10$^5$ at high altitudes where pressure is $\sim$~0.02 bar or less (see the fully-saturated H$_2$O$_2$ concentration shown in Fig.\ref{fig:tpfp}~b). 
Condensation of H$_2$O$_2$ in the high altitude atmosphere is discussed in Sec.~\ref{ssec:c}.}

\section{Discussion} \label{sec:disc}

\subsection{H$_2$O$_2$ in a wet and oxidized atmosphere} \label{ssec:p}
H$_2$O$_2$ is much more abundant in a wet and oxidized atmosphere, though the concentration of H$_2$O$_2$ in the current dry Martian atmosphere is about 10~ppb \citep{Encrenaz+2004}. 
Although chemical models suggest that the concentration of H$_2$O$_2$ reaches at most 0.1 ppm in dry atmospheres \citep{Parkinson+1972,Gao+2015}, a wet and oxidized atmosphere which is suitable for the formation of H$_2$O$_2$ would contain  H$_2$O$_2$ in a concentration higher than 0.1 ppm.  
This is because  H$_2$O$_2$ is produced through the chemical reactions of HO$_x$ gas species such as H, OH and HO$_2$ which originate from H$_2$O. 
Also, the abundance of H$_2$O$_2$ would be higher in an oxidized atmosphere because such an atmosphere inhibits the regeneration of H$_2$O from HO$_x$ and enhances the production of H$_2$O$_2$.

In a wet and oxidized Martian atmosphere, the photolysis of H$_2$O$_2$ is considered to be an effective pathway to regenerate CO$_2$ \citep{Yung+1999}.  
This regeneration is necessary because CO$_2$ is destroyed by far-UV irradiation ($\lambda\leq$227.5 nm) from the Sun via; 
\begin{align}
\mathrm{CO}_2+h\nu \to \mathrm{CO} + \mathrm{O}. \tag{R1}
\end{align}
Indeed CO$_2$ regeneration is required to maintain the CO$_2$ atmosphere over geological timescales.  
In a wet atmosphere, H$_2$O$_2$ can be sufficiently produced as an intermediate product through the following catalytic cycle:
\begin{align}
2(\mathrm{H} + \mathrm{O}_2 + \mathrm{M} &\to \mathrm{HO}_2 + \mathrm{M}),  \tag{R2} \\
2\mathrm{HO}_2  &\to \mathrm{H_2O_2} + \mathrm{O}_2,  \tag{R3} \\
\mathrm{H_2O_2} + h\nu  &\to 2\mathrm{OH},  \tag{R4} \\
2(\mathrm{OH} + \mathrm{CO}  &\to \mathrm{CO}_2 + \mathrm{H}),  \tag{R5} \\
------&------- \nonumber \\
2\mathrm{CO} + \mathrm{O}_2  &\to 2\mathrm{CO}_2.  \tag{S1}
\end{align}
Meanwhile, although a thick and dry CO$_2$-rich atmosphere is unstable \citep{Zahnle+2008}, in a wet and oxidized atmosphere of early Mars, CO$_2$ could have been stabilized by S1 (=R2+R3+R4+R5) even if the atmosphere was thick.

We estimate the H$_2$O$_2$ abundance in a warm/wet and oxidized CO$_2$ atmosphere by assuming that S1 is the cycle most responsible for the regeneration of CO$_2$ against loses due to R1.
We assume that the bulk CO$_2$ abundance in the atmosphere is in balance between its photo-dissociation flux (R1), and twice the H$_2$O$_2$ photo-dissociation flux (R4) that produces OH for oxidizing CO.
Also, it is assumed that there is no optical shielding effect for photons with wavelengths longer than the shielded wavelength, $\lambda_\mathrm{sh}$, but there is complete shielding for all other UV photons to H$_2$O$_2$, for simplicity. Then, the vertical column density of H$_2$O$_2$, $\Sigma_\mathrm{H_2O_2}$, can be written as;
\begin{equation}
\Sigma_\mathrm{H_2O_2}= \frac{\int_{\lambda\leq227.5\mathrm{nm}} \hat{F_\lambda}d\lambda}
{2\int_{\lambda_\mathrm{sh}}^{\lambda_\mathrm{th}} \hat{F_\lambda} \sigma_{\mathrm{diss}, \lambda} d\lambda},
\label{eq:am_h2o2}
\end{equation}
where $\hat{F_\lambda}$ is solar photon flux, and $\sigma_{\mathrm{diss}, \lambda}$ and $\lambda_\mathrm{th}$ are the photo-dissociation cross section and the threshold wavelength for a photon to effectively dissociate H$_2$O$_2$, respectively.
Owing to the low bonding energy of H$_2$O$_2$ 
($\sim$ 50 kcal/mol $\sim$ 570~nm; \citealt[][]{Bach+1996}), the photo-dissociation is caused not only by UV but also by visible light photons.
Therefore, H$_2$O$_2$ is not completely shielded from stellar irradiation by H$_2$O, O$_2$ and CO$_2$ \citep{Yung+1999}.
On the other hand, a developed O$_3$ layer may shield solar photons with wavelengths $\lesssim$ 300 nm, as displayed on Earth today.
Here we use $\lambda_\mathrm{sh}=227.5$~nm and 300~nm as fiducial values of a shielded wavelength.

The dissociation cross section of H$_2$O$_2$ has been measured only for photon wavelengths in the range $\leq$ 410~nm \citep{Kahan+2012} because of the technical problem of measuring small absorption cross sections.
Hence, we use $\lambda_\mathrm{th}=410$~nm as a fiducial value of the threshold wavelength.
Note that, according to \citet{Kahan+2012}, the photolysis of H$_2$O$_2$ mainly occurs at photon wavelengths shorter than 350~nm.
Therefore, inputing $\lambda_\mathrm{sh}=227.5$~nm, the measured cross section
with $\lambda_\mathrm{th}=$ 410~nm \citep{Lin+1978,Kahan+2012} and the solar spectral 
irradiance at 4~Ga
 developed by combining the observed spectrum from the Sun with those of solar-type stars at different ages \citep{Claire+2012}
 in Eq.~(\ref{eq:am_h2o2}), we find $\Sigma_\mathrm{H_2O_2}\sim8\times10^{17}$~cm$^{-2}$.
When we substitute $\lambda_\mathrm{sh}=300$~nm into Eq.~(\ref{eq:am_h2o2}), we find
$\Sigma_\mathrm{H_2O_2}\sim5\times10^{18}$~cm$^{-2}$. 
These values change only 10~\% if the solar spectral irradiance at 3.5~Ga is used instead.

The column densities estimated here are significantly larger than the current typical value of $\sim2\times10^{15}$~cm$^{-2}$, which corresponds to  10 ppb at 6 mbar, in the present-day Martian atmosphere.
These large column densities produce optical depth over wavenumbers $\nu=100$--500~cm$^{-1}$ of $\tau_\nu=0.003$--0.6 for $\lambda_\mathrm{sh}=227.5$~nm and $\tau_\nu=0.02$--4 for $\lambda_\mathrm{sh}=300$~nm, assuming a far-IR absorption cross section of  H$_2$O$_2$, $\sigma_{\nu, H_2O_2}=0.04$--$8 \times10^{-19}$~cm$^2$, which is shown in Fig.~\ref{fig:cross}.
Thus, if the other gases such as O$_3$ sufficiently can reduce the photolysis of H$_2$O$_2$, then the amount of H$_2$O$_2$ in the atmosphere would be large enough to warm the planetary surface.
Note that, the column density of H$_2$O$_2$ estimated by Eq.~(\ref{eq:am_h2o2}) is just a typical value when S1 is the cycle most responsible for the stabilization of CO$_2$, while this value could be increased if the self-shielding effect was taken into account in Eq.~(\ref{eq:am_h2o2}).
This is because we impose the restriction that only the OH produced by the photolysis of H$_2$O$_2$ is used to oxidize CO via R5, but all other reactions which produce and remove OH are ignored.
Also, the other process potentially affecting the concentration of H$_2$O$_2$ is discussed in Sec~\ref{ssec:op}.

\subsection{Condensation of H$_2$O$_2$} \label{ssec:c}
It is likely that H$_2$O$_2$ in a warm and wet atmosphere of early Mars is super-saturated  \added{at high altitudes} because the timescale for condensation is likely longer than that for photochemical production.  
As described later, a timescale for condensation would be much longer than that that governing the production and photo-dissociation of H$_2$O$_2$, which was shown by \citet{Nair+1994}, who used a photochemical model, to be of order several hours.

The condensation time can be estimated by assuming that H$_2$O$_2$ condenses as soon as it collides with condensation nuclei, namely;
\begin{eqnarray}
   &{\tau_\mathrm{cond}}& = \left(4\pi r^2 N_{\mathrm{ccn}} \rho v_{T} \right)^{-1}, \label{eq:ct} \\
&& \sim 50 \mathrm{\ hours} 
\times  
\\
 && \left(\frac{N_{\mathrm{ccn}}}{10^5\mathrm{kg}^{-1}}\right)^{-1}
 \left(\frac{P}{0.01\mathrm{bar}}\right)^{-1}
 \left(\frac{T}{200\mathrm{K}}\right)^{\frac{1}{2}}
 \left(\frac{r}{1\mathrm{\mu m}}\right)^{-2}_,
 \nonumber
\end{eqnarray}
where $r$ and $N_{\mathrm{ccn}}$ are the size and concentration of the condensation nuclei, respectively, $\rho$ is the atmospheric mass density and $v_T$ is the thermal velocity of the gas.
The timescale for condensation is longer at higher altitudes because the nuclei concentration decreases with increasing altitude.
Note that, the condensation timescale is underestimated in an atmospheric region with a mean free path smaller than the size of the nuclei (i.e., a dense region) because the diffusive motion of the gas around  the nuclei delays the timescale \citep[see][for the diffusive case]{Lohmann+2016}.

\begin{figure}[t]
  \begin{center} 
  \includegraphics[width=\columnwidth]{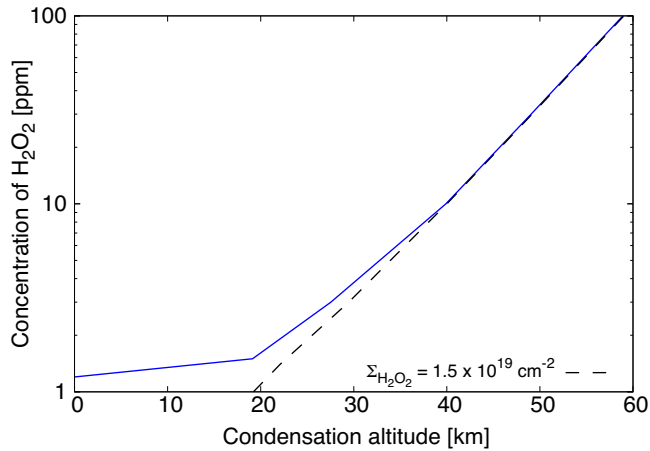}
  \\
  \caption{
The concentration of H$_2$O$_2$ necessary to warm the planetary surface to 273~K in the 2--bar atmosphere as a function of the condensation altitude (see the text for the definitions of each term).
The dashed line represents the column density of H$_2$O$_2$ with a constant concentration to reach $\Sigma_\mathrm{H_2O_2}=$1.5$\times$$10^{19}$~cm$^{-2}$.
}
  \label{fig:h2o2_sat}
  \end{center}
\end{figure}

Achieving sufficient warming is possible even if H$_2$O$_2$ condenses at lower altitudes due to the subsequent shorter condensation times. 
Fig.~\ref{fig:h2o2_sat} shows the minimum H$_2$O$_2$ concentration necessary for maintaining a surface temperature of at least  273~K in a 2-bar atmosphere as a function of a condensation altitude.  
The condensation altitude stands for an altitude above which the H$_2$O$_2$ concentration is constant and below which all the H$_2$O$_2$ gas is virtually removed by rainout through condensation.
To warm the surface environment, the required concentration of H$_2$O$_2$ needs to be about 2 ppm when the condensation altitude is no higher than about 20 km.
The 2~ppm of H$_2$O$_2$ in the upper atmosphere is comparable to 1.5$\times$10$^{19}$~cm$^{-2}$ which is also comparable to the H$_2$O$_2$ column density necessary to stabilize the CO$_2$ atmosphere (Sec~\ref{ssec:p}).  

\added{
Recent measurements of water vapor in current Martian upper atmosphere indicate that the supersaturation of water vapor ranges 1--10$^2$ \citep{Maltagliati+2011,Fedorova+2020},  much smaller than $10^5$ that is the supersaturation of 1 ppm H$_2$O$_2$.
However, it should be noted that the mechanism of supersaturation is completely different.
On Mars today, the supersaturation of water vapor is generated by the transport of water vapor from the higher temperature region to the lower temperature region.
On the other hand, the supersaturation of H$_2$O$_2$ would be generated by in situ photochemical production with a timescale of several hours.
It is likely that the H$_2$O$_2$ production generates higher supersaturation than that generated by transport, though further studies are needed to evaluate the degree of supersaturation, quantitatively.

Warming the planetary surface is still possible if the supersaturation is suppressed to $10^3$ level.
In the 2-bar atmosphere with a surface temperature of 273~K, 1~ppm of H$_2$O$_2$ makes the outgoing radiation to 68.8~W/m$^2$, though the supersaturation is as high as ~10$^5$ at higher altitude (see Sec.~\ref{sec:res}).
When the supersaturation has an upper limit of $S_c = 10^4$, $10^3$, or $10^2$, the outgoing radiation in our model are 70.5~W/m$^2$, 74.5~W/m$^2$ or 79.5~W/m$^2$, respectively.
The upper limit of supersaturation has a rather little influence on the outgoing radiation since the photosphere for the wavenumbers of $\geq$ 500~cm$^{-1}$ is around 0.3~bar level where the supersaturation is $\sim10^3$.
Also, in the case of the 3-bar atmosphere with 1~ppm of H$_2$O$_2$, $S_c>2\times 10^3$ is required to warm the surface temperature above 273~K.
}

The condensation timescale will not be significantly changed if the dilution effect of H$_2$O$_2$ in an H$_2$O solution is taken into account.  
When the temperature is above $\sim220$~K, an H$_2$O--H$_2$O$_2$ solution can exist, and then the saturation vapor pressure of H$_2$O$_2$ will be lowered relative to that of pure H$_2$O$_2$ \citep{Foley+1951b,Manatt+2004}.
However, the temperatures in the photosphere for photons with wavenumbers in the range $\geq$ 500~cm$^{-1}$ are lower than 220~K in thick and warm atmospheres (Fig.~\ref{fig:tpfp}).  
So, it is likely that aqueous solutions would be frozen in the upper atmosphere where the concentration of H$_2$O$_2$ has the greatest influence on the surface temperature.  
Therefore, the dilution effect of H$_2$O$_2$ in an H$_2$O solution would little affect the surface temperature.

\added{
H$_2$O$_2$ clouds would affect the surface temperature if they are formed. In general, low-altitude clouds can cool the planetary surface and high-altitude clouds can warm the surface \citep[][]{RK+2017}. It would be important to discuss the detail of the H$_2$O$_2$ cloud radiative forcing but we have left it to the future studies since there is no public refractive index data of H$_2$O$_2$ particles (see Refractive index database\footnote{\url{https://refractiveindex.info}}).
}

\subsection{Other processes possibly affecting H$_2$O$_2$ concentration}\label{ssec:op}
The atmospheric concentration of H$_2$O$_2$ can also be affected by several processes such as dissolution into water droplets, dry deposition and photochemical reactions with volcanic and reactive species (e.g., SO$_2$ and NO$_x$) \citep{Vione+2003}.

Although H$_2$O$_2$ is a minor species with at most 3.5 ppb level in the Earth's atmosphere, which is mainly due to the dissolution of gaseous H$_2$O$_2$ into water droplets, where SO$_2$ enhances the dissolution rate \citep{Vione+2003}, it might not be a minor species on early Mars during the era in which the observed Mn-oxide-rich rocks at Gale crater would have precipitated out.
Since the temperatures at high altitudes in the early Martian atmosphere would be so low that H$_2$O would freeze, its non-dissolution into water droplets would not deplete H$_2$O$_2$. Meanwhile, at lower altitude regions, H$_2$O$_2$ would dissolve into water droplets, and precipitation would remove it.

\added{
If the abundant sulfur-bearing gases were supplied to early Martian atmosphere, SO$_2$ might have destroyed H$_2$O$_2$ \citep[e.g.,][]{Spracklen+2005,Galeazzo+2018}.
Although Mn and S are not correlated in the rocks found at Gale crater \citep{Lanza+2016}, sulfur deposits are present in large amounts all across the planet and they date to about the same period as Gale crater \citep[e.g.,][]{Bibring+2005,Gendrin+2005}.  
}

In Earth's atmosphere, dry deposition is another removal process of atmospheric H$_2$O$_2$ at lower altitudes.
Atmospheric H$_2$O$_2$ of early Mars would be vertically transported by eddy diffusion to the surface, whereby dry deposition and precipitation remove it.
For the current Martian atmosphere at altitudes lower than 40~km, the scale height is $H \sim 10$ km and the vertical eddy diffusion coefficient is $K_{\mathrm{ed}} \leq 10^7$cm$^2$/s \citep{Nair+1994}; hence the diffusion timescale is $H^2/K_{\mathrm{ed}} \geq 1$~days.
Since the timescale of H$_2$O$_2$ photochemical reactions is less than a day \citep{Nair+1994, Zahnle+2008}, the atmospheric concentration of H$_2$O$_2$ at high altitudes is likely controlled by photochemical reactions. 

The actual eddy diffusion coefficient and dry deposition timescale on early Mars would depend on turbulence/large-scale-winds and the compositions/oxidations states of the surface rocks, respectively.
As such, a more detailed examination requires that photochemical calculations be done alongside those of the atmospheric thermal structure, which will be the focus of a future study.

\subsection{Oxidized early Martian environment} \label{ssec;o}

An early surface environment warmed by the greenhouse effect of H$_2$O$_2$ (Sec. \ref{ssec:p}) is consistent with the global, highly oxidized conditions implied by the high Mn materials found on the Martian surface by the Curiosity rover in Gale crater and by the Opportunity rover in Endeavour crater \citep{Lanza+2016,Arvidson+2016}.

The redox state of early Martian atmosphere is likely controlled by the escape of atmospheric components into space.  
In the early Martian atmosphere, 
UV radiation from the young Sun would have enhanced hydrogen escape
and effectively oxidized the atmosphere and the surface environment.  
In addition to hydrogen escape, the escape of atomic carbon might also have contributed to the oxidation of the early Martian atmosphere because its escape flux would not be limited by diffusion in a CO$_2$-rich atmosphere (N. Terada, private communications).  
Further studies are required to determine the redox state of the early Martian atmosphere, which could also be affected by the supply of reduced gases (e.g., CO and H$_2$) through volcanic degassing, oxygen escape, and oxygen uptake through weathering of the planetary surface \citep{Zahnle+2008,Wetzel+2013,Batalha+2015}.  

It is interesting to note that H$_2$O$_2$ might be able to warm a frozen planet and melt water ice.  
\citet{Liang+2006} demonstrated that a weak hydrological cycle coupled with photochemical reactions could give rise to a sustained production of H$_2$O$_2$ during long and severe glacial intervals.  
Although an icy surface has a high albedo, the surface temperature can be warmed to temperatures above 273~K by a 4 and 15 ppm levels of H$_2$O$_2$ in a 2~bar atmosphere when the planetary albedo is assumed to be $\leq$ 0.45 and $\leq$ 0.5, respectively, as demonstrated by our model.  

It has also been suggested that H$_2$O$_2$ deposited on the planetary surface could be stored in the ice during the time of a  global snowball episode \citep{Liang+2006}.  
If early Mars was once a snowball, and a large amount of H$_2$O$_2$ was stored in the ice, it would be released into the atmosphere upon melting caused by any mechanism, such as meteor impacts, volcanic emissions, or obliquity changes \citep[e.g.,][references therein]{Wordsworth2016}. 
The release of abundant H$_2$O$_2$ would cause not only a global oxidation event but also enhance greenhouse warming.  
If so, there might be geological evidence that oxidation and warming occurred simultaneously in the aftermath of a snowball Mars.

\section{Summary and Conclusion} \label{sec:conc}
We investigated the possible impact of H$_2$O$_2$ as an additional greenhouse gas in a CO$_2$-dominant atmosphere using a one-dimensional atmospheric model.  
Because the timescale for condensation is longer at higher altitudes (subsection~\ref{ssec:c}), photochemically produced H$_2$O$_2$ would likely be supersaturated in the upper atmosphere.  
We found that a reasonable amount of H$_2$O$_2$ in the upper atmosphere effectively cuts off the outgoing planetary radiation in the far-infrared and warms the planetary surface to a temperature hot enough to retain liquid water (Section~\ref{sec:res}).

Our results demonstrated that a warm and wet surface environment is compatible with an oxidized atmosphere on early Mars.  
The coexistence of liquid water and an oxidized atmosphere on early Mars has been suggested by the recent discovery of a high level of Mn in some Martian rocks \citep{Lanza+2016,Arvidson+2016}.
Our results also indicated a key relationship between the redox state of the atmosphere and the surface temperature on early Mars, where the co-evolution of these factors may govern the surface environment over geological time scales. This important phenomenon will be the subject of future work, which will aim to understand the surface environment under an oxidized atmosphere on early Mars.

\acknowledgements
We thank Ramses Ramirez for sharing their albedo data of CO$_2$ atmospheres with us.
This work was supported by MEXT/JSPS KAKENHI Grants Numbers 17H06457, 18K03719, and 19H01947 and by
NINS Astrobiology Center Project Grant Numbers AB311025.

\bibliography{ref_mars}{}
\bibliographystyle{aasjournal}

\end{document}